\documentclass[prb,aps,twocolumn,amsmath,amssymb,superscriptaddress]{revtex4-2}

\usepackage{times}
\usepackage{textcomp}
\usepackage{graphicx}
\usepackage[caption=false]{subfig}
\usepackage{bm}
\allowdisplaybreaks 
\usepackage{braket}
\usepackage{lipsum}
\usepackage{amsmath}

\newcommand{\tmmathbf}[1]{\ensuremath{\boldsymbol{#1}}}

\makeatletter

\newcommand{\Rmnum}[1]{\expandafter\@slowromancap\romannumeral #1@}
\makeatother

\usepackage[breaklinks]{hyperref}
\hypersetup{colorlinks=true, linkcolor=blue, citecolor=blue, filecolor=blue, urlcolor=blue}

\begin{document}

\title{Logical Majorana zero modes in a nanowire network}

\author{Sayandip Dhara}
    \affiliation{Department of Physics, University of Central Florida, Orlando, Florida, 32816, USA}

\author{Garry Goldstein}
    \affiliation{Physics Department, Boston University, Boston, Massachusetts 02215, USA}

\author{Claudio Chamon}
    \affiliation{Physics Department, Boston University, Boston, Massachusetts 02215, USA}    

\author{Eduardo R. Mucciolo}
    \affiliation{Department of Physics, University of Central Florida, Orlando, Florida, 32816, USA}

\date{\today} 

\begin{abstract}
  We present a scheme to use physical Majorana quasi-zero modes at
  each junction of a two-dimensional nanowire network to build a
  logical Majorana zero mode, the location of which is controllable
  through gate voltages. The wire-network is a way to realize a
  proposal by Yang {\it et
    al}. \cite{yangHierarchicalMajoranasProgrammable2019a} to imprint
  a Kekul\'{e} vortex pattern on a honeycomb lattice via gate
  voltages. We show that a specific type of junction -- other than a
  naive Y- or T-junction -- is needed to realize, without breaking
  time-reversal symmetry, an artificial ``graphene'' system with
  Majorana fermions instead of complex ones at each site. The junction
  we propose (i) traps exactly one physical Majorana (quasi-)zero mode
  at each site of either a brick wall or honeycomb lattice and (ii)
  allows this mode to hybridize with all three neighboring
  sites. Using a lattice of these junctions and starting from an
  electronic, tight-binding model for the wires, we imprint the
  voltage patterns corresponding to the Kekul\'{e} vortex and observe
  the emergence of the logical Majorana zero mode at the vortex
  core. We also provide the range of parameters where this excitation
  could be realized experimentally.
\end{abstract}


\maketitle

\section{Introduction}
\label{sec:introduction}

Quantum computation has the potential to solve problems that are
intractable in current, classical computers. This potential will be
realized when large-scale, fault-tolerant quantum computers become
available. In the quest for such machines, two distinct approaches
stand out: to increase the number of physical qubits, so that logical
qubits can be created and quantum error correction can be applied;
\cite{dennisTopologicalQuantumMemory2002,bravyiQuantumCodesLattice1998,Fowler};
or, alternatively, to build qubits that are intrinsically
decoherence-resistant due to topological protection
\cite{kitaevFaulttolerantQuantumComputation2003a}. In this paper, we
explore the latter but infuse it with elements of the former.

Topological qubits \cite{nayakNonAbelianAnyonsTopological2008,
  kitaevAnyonsExactlySolved2006a, kitaevUnpairedMajoranaFermions2001a,
  fuSuperconductingProximityEffect2008} based on Majorana zero modes
are an example where the qubit has protection against local noise
because the information is encoded non-locally, shared between distant
localized zero modes. There have been a number of proposals
\cite{aliceaNonAbelianStatisticsTopological2011a,
  fujimotoTopologicalOrderNonAbelian2008,
  lutchynMajoranaFermionsTopological2010a,
  aliceaMajoranaFermionsTunable2010, oregHelicalLiquidsMajorana2010a}
in the last decade on how to realize phases with Majorana zero modes
at the endpoints of one-dimensional (1D) wires obtained by interfacing
superconducting and semiconducting materials. There has been enormous
progress on engineering these material interfaces, as well as on
characterizing the properties of the 1D wires and detecting zero-bias
peaks at their extremities~\cite{mourikSignaturesMajoranaFermions2012,
  lutchynSearchMajoranaFermions2011,
  sauGenericNewPlatform2010}. However, there are still ongoing debates
on the exact nature of the zero modes in experimental systems
\cite{aghaeeInAsAlHybridDevices2022}, as well as on the proper
protocol to separate topological zero-mode states from non-topological
Andreev bound states
\cite{liuAndreevBoundStates2017a,saulsAndreevBoundStates2018}.

\begin{figure}[h]
  \raisebox{0.0\height}{\includegraphics[width=8cm,height=6cm]{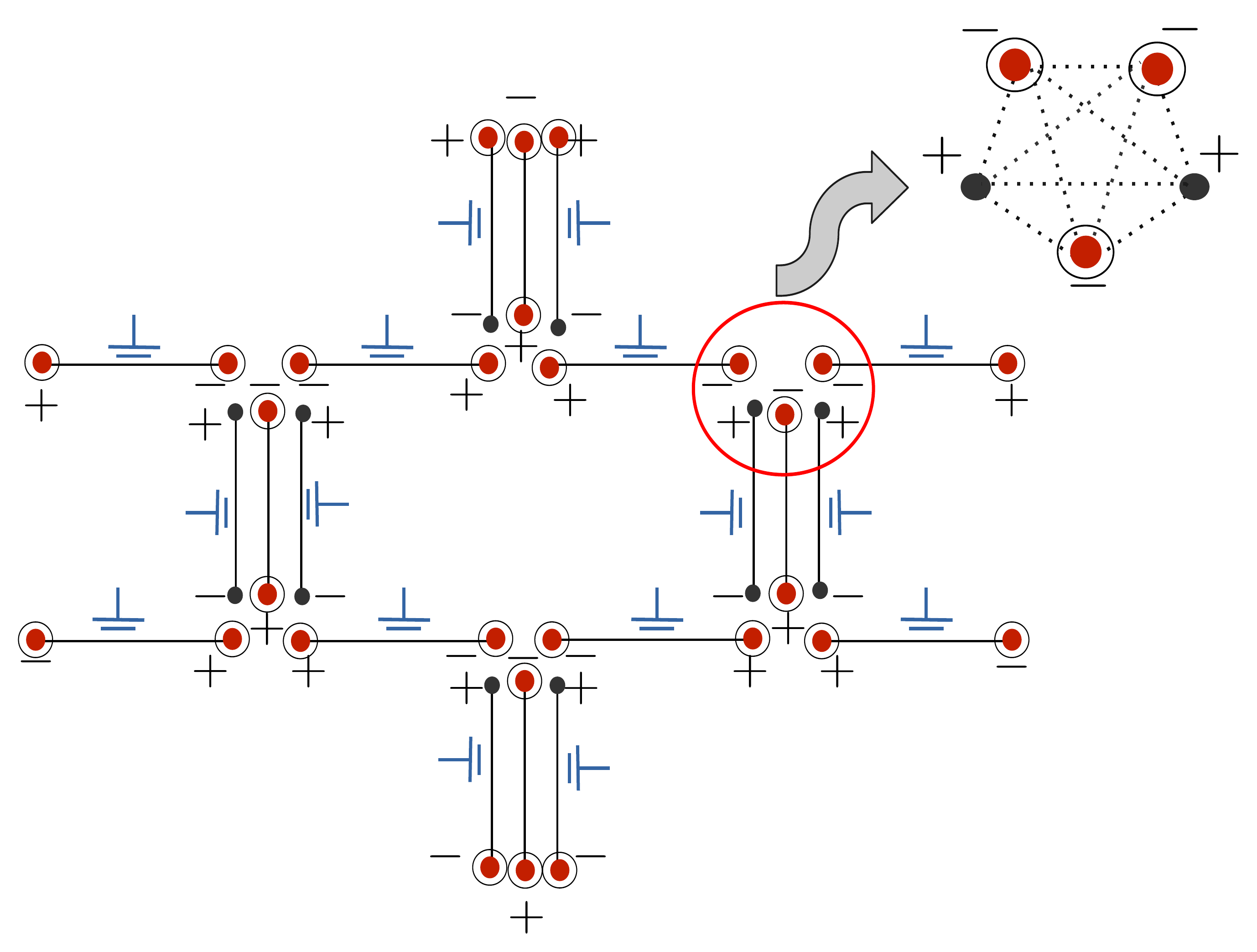}}
  \caption{A unit cell of the brickwall network. Every junction
    consists of five nanowires and the polarity ($\pm$ signs)
    correspond to whether the zero mode at a given endpoint of a wire
    is even or odd under time-reversal symmetry. Notice that the five
    wires are arranged in such a way that two wires ``sister'' a
    (third) central wire in one of the legs of the brickwall
    lattice. The presence of a nonzero Majorana zero mode amplitude at
    a nanowire end is indicated by a circular line around a solid
    circle. The polarities on the nanowires are such that there is
    only one zero mode per lattice site, and that this mode hybridizes
    will all the three neighboring sites. The red circle shows a
    zoom-in of the interwire couplings at a junction.}
  \label{fig:brickwall}
\end{figure}

In this paper, we use 1D wires as building blocks to construct {\it
  logical} Majorana zero modes on a two-dimensional (2D) wire
network. The location of these logical Majorana zero modes is
controlled by applied gate voltages on the wires in the network. The
network that we propose builds on that by Yang {\it et
  al}. \cite{yangHierarchicalMajoranasProgrammable2019a}, where a
hierarchical framework was used to build a logical Majorana zero mode
in a two-dimensional (2D) honeycomb network with links consisting of
1D nanowires. The proposal contains three steps. In the first step,
each finite-length nanowire in the network is brought to a regime
where there is a single Majorana zero mode at each nanowire end. At
the vertices of the network, where three nanowires meet, the Majorana
zero modes hybridize. As a result, two of the zero modes are gapped
out and only one survives, leaving one Majorana zero mode per
vertex. In the second step, by tuning the gate voltages on each wire,
these surviving zero modes are made to weakly overlap, creating a
band. In the third and final step, a Kekul\'{e} vortex modulation
\cite{chamonSolitonsCarbonNanotubes2000,
  houElectronFractionalizationTwoDimensional2007,
  ryuMassesGraphenelikeTwodimensional2009} of gate voltages is
employed to open a spectral gap everywhere but on the vortex, which
binds a topological zero mode.

Two conditions are essential for obtaining a ``Majorana graphene''
system on which to build the logical Majorana zero mode at the vortex
core: (i) having a single Majorana mode on each site of the honeycomb
lattice, and (ii) hybridizing this mode with all three neighboring
sites. In Ref.~\cite{yangHierarchicalMajoranasProgrammable2019a} this
was achieved by a fixed pattern of breaking time-reversal symmetry
(TRS) at the junction of the three wires connecting at each
site. While breaking TRS is not a problem in practice, breaking it in
a prescribed pattern on the lattice -- in opposite ways in the two
sublattices of the honeycomb lattice -- could be difficult to
achieve. In this paper, we provide a concrete realization of the
``Majorana graphene'' system of
Ref.~\cite{yangHierarchicalMajoranasProgrammable2019a} that requires
no breaking of TRS. Here we show that it is impossible, in systems
with TRS, to satisfy both conditions (i) and (ii) with junctions of
three wires, such as Y- or T-junctions. To satisfy both conditions
requires the use of junctions with five wires, where two wires
``sister'' a third central wire in one of the legs of the Y- or
T-junctions, as depicted in Fig.~\ref{fig:brickwall}.

For systems with TRS, one can define two opposite polarities for the
zero modes at the endpoints of a wire, corresponding to whether the
zero modes are even ($+$) or odd ($-$) under reversal of time. (Which
end is assigned $+$ or $-$ depends on the couplings in the
Hamiltonian, for example, the sign of the superconductor order
parameter, as we discuss in the paper.) The number of zero modes at a
junction of $n=n_++n_-$ wires, with $n_+$ of positive and $n_-$ of
negative polarity, is given by the integer-valued index $\nu = |n_+ -
n_-|$. (The superconducting 1D wires with TRS are in symmetry class
BDI~\cite{Schnyder_et_al-2008}, which is indexed by a topological
invariant $\nu \in \mathbb Z$; interactions, which we do not include,
break the classification down to $\mathbb
Z_8$~\cite{Fidkowski-Kitaev-2010}.) Satisfying the condition (i) above
is thus possible with three wires if two have one polarity, and one
the other, so that $\nu = |2-1|=1$. However, as we discuss in this
paper, the wave function for the zero mode has amplitude only on the
{\it majority} wires, i.e., only on $\rho=\max(n_+,n_-)$ wires (far
away from the junction). In the case of junctions of three wires
above, the wave function on one site would leak to only
$\rho=\max(2,1)=2$ out of the three neighbors, yielding a system of
decoupled 1D Majorana chains. Therefore it is condition (ii) that
poses an obstruction to constructing a 2D system of ``Majorana
graphene'' with Y- or T-junctions. With five-wire junctions as
depicted in Fig.~\ref{fig:brickwall} we solve the problem with $n_\pm
= 3$ and $n_\mp=2$ on the two sublattices of the brickwall lattice,
so that $\nu=|3-2|=1$ and $\max(3,2)=3$, which satisfy both conditions
(i) and (ii).

Using the geometry of Fig.~\ref{fig:brickwall}, we then construct an
electronic tight-binding model of the nanowires and junctions that
realize an effective honeycomb lattice of Majorana quasi-zero
modes. We implement a Kekul\'{e} modulation on the potential of the
tight-binding nanowires that opens a bulk spectrum gap, and create
vortex that binds a single zero mode to a particular
location. Numerical simulations confirm the exact location of the
vortex and the ability to move the zero mode around the lattice by a
simple change in the gate voltage modulation. To show the presence of
a logical Majorana quasi-zero mode, we compute the exact local density
of states in the nanowire network in the presence of a vortex. We use
our results to estimate the parameters required for an experimental
realization and discuss whether the detection of the logical Majorana
mode is possible. One of the most important takeaways of this approach
of building a logical Majorana zero mode from a collection of physical
Majorana quasi-zero modes is that we can afford less stringent
conditions on the physical Majoranas at the nanowire level. The
individual nanowires do not need to provide two highly localized zero
modes at both ends of the wire; in fact, we exploit the opposite, that
in experiments, the nanowire zero modes likely hybridize.

The paper is organized as follows. In Sec.~\ref{sec:indices} we
present the design rationale for the wire network in
Fig.~\ref{fig:brickwall}. In Sec. \ref{sec:wires_and_junctions}, we
discuss a simple model of the 1D quantum wires, and present numerical
studies with junctions of three and five wires that support and
justify our choice of wire network in Fig.~\ref{fig:brickwall}, which
we then study in Sec. \ref{sec:network}, where we introduce the
Kekul\'{e} modulation of the network to gap the zero modes in the bulk
and then demonstrate the gate voltage modulation required to imprint a
vortex inside the lattice. In Sec. \ref{sec:parameters} we describe
the realistic parameters necessary for the experiments. We conclude in
Sec. \ref{sec:conclusions} with a summary and a discussion of open
questions.

\section{The nanowire network design rationale}
\label{sec:indices}

In this section, we justify the design of the nanowire network in
Fig.~\ref{fig:brickwall}. Our goal is to obtain an artificial
``Majorana graphene'' platform, i.e., a system in which single
Majorana (quasi-)zero mode sits on the sites of a brickwall or a
honeycomb lattice and hybridizes with the three neighboring sites. This
platform shares many of the features of graphene but with Majorana
(not complex) fermions on the sites. The programmable hoppings (via
gate voltages) allow us to imprint Kekul\'e vortices in the
dimerization pattern, thereby trapping logical Majorana zero modes.

The conditions to realize the ``Majorana graphene'' platform, namely,
(i) that a single (quasi-)zero mode sits on each site, and (ii) that
these modes hybridize with the three neighbors, and are connected to two
indices that we discuss in this section. Here we shall focus solely on
an effective model of Majorana end modes on the nanowires, without
diving into any microscopic model of the nanowires or the junctions;
that discussion is reserved for the subsequent sections.

Let us start from a single nano wire in a phase in which two Majorana
zero modes sit at the wire endpoints, which we label
$\gamma_\pm$. Under the TRS operation ${\cal T}$, one of these modes
is even and the other is odd:
\begin{align}
  {\cal T}\;\gamma_{\pm}\;{\cal T}^{-1}
  =
  \pm\,\gamma_{\pm}
  \;.
\end{align}
The $\pm$ sign associated with the parity of the endpoint zero modes can
be thought of as a polarity for that endpoint. Notice that the only
possible coupling that can be added to the effective model in which
there are only the Majorana endpoints left in the wire is
$H=i\,\gamma_+\,\gamma_-$, which is non-local and gaps the wire. That
$\gamma_\pm$ have opposite polarity is needed for this $H$ to be both
Hermitian and respect TRS.

\begin{figure}[h]
  \raisebox{0.0\height}{\includegraphics[width=5cm]{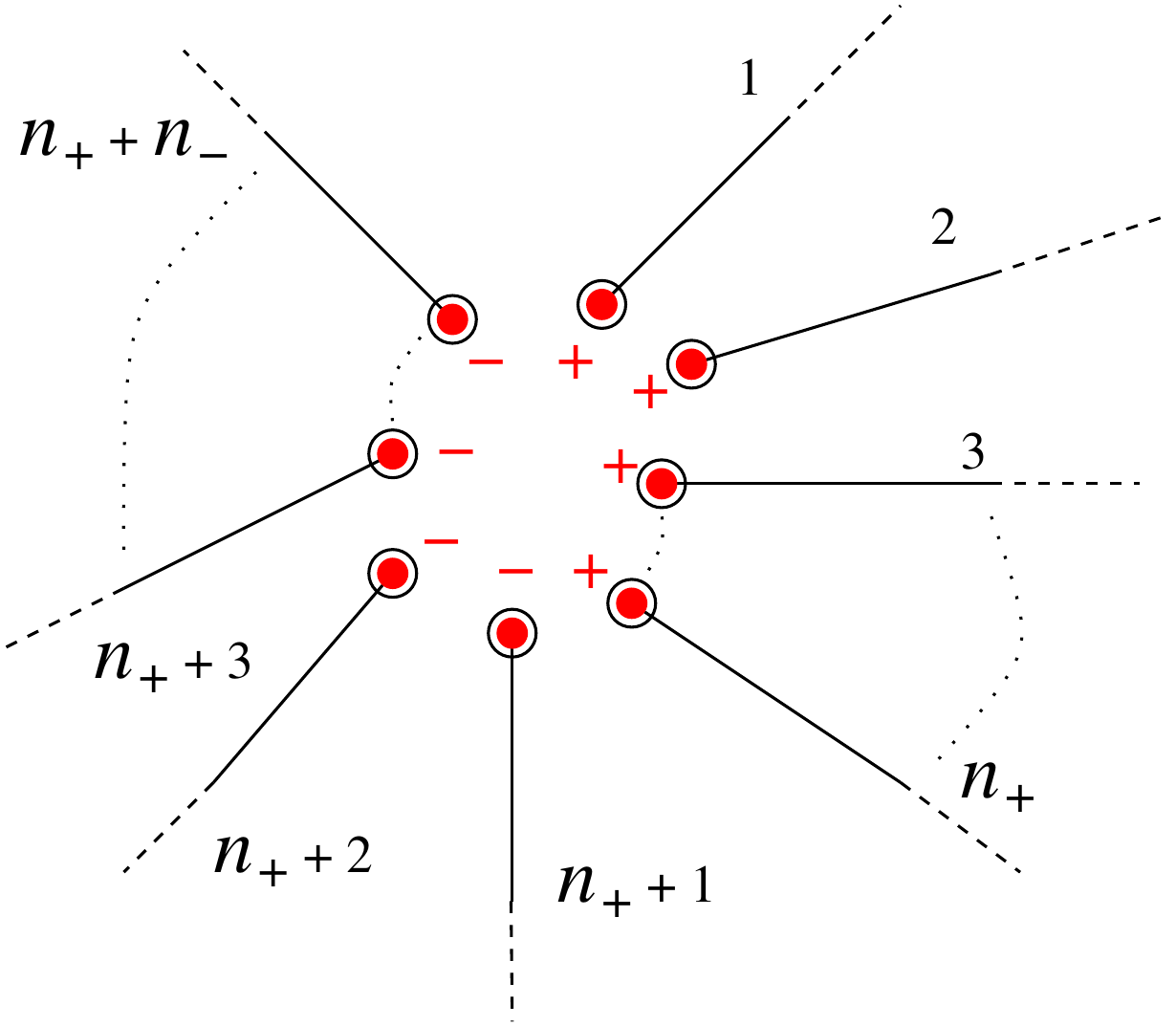}}
  \caption{Schematic illusrtration of a multi-nanowire junction. $n_+$
    ($n_-$) nanowires have positive (negative) polarity at the inner
    endpoints, which are represented by red circles.}
    \label{fig:n-junction-polarity}
\end{figure}

Consider now a junction where endpoints of multiple wires come
together, of which $n_+$ have positive polarity and $n_-$ have
negative polarity, as shown in Fig.~\ref{fig:n-junction-polarity}. The
tunneling Hamiltonian $H_{\rm junction}$ is quadratic in
$\gamma_{a+}$, $a=1,\dots, n_+$ and $\gamma_{b-}$, $b=1,\dots,
n_-$. Moreover, to respect TRS, $H_{\rm junction}$ must only couple
even with odd, or $+$ with $-$, modes, i.e., the Hamiltonian must have
the form
\begin{align}
H_{\rm junction} = i\;\sum_{a=1}^{n_+} \sum_{b=1}^{n_-}
  \gamma_{a+}\, \Gamma_{ab}\, \gamma_{b-}
  + H.c.
  \;,
\end{align}
where $\Gamma_{ab}\in \mathbb R$. The eigenmodes can be obtained from
the spectrum of the $(n_++n_-)\times (n_++n_-)$ matrix
\begin{align}
  h = i
  \left[
  \begin{array}{c|c}
    0_{n_+\times n_+}&\Gamma_{n_+ \times n_-}\\
    \hline
    -\Gamma^\top_{n_-\times n_+}&0_{n_- \times n_-}
  \end{array} \right]
  \;,
  \label{eq:h_junc}
\end{align}
The number of zero modes of this block matrix (in the generic case) is
given by the index
\begin{align}
  \nu = |n_+-n_-|
  \;.
\end{align}
One can trace this index to the fact that the system of 1D wires with
TRS symmetry belongs to class BDI in the classification of topological
insulators and superconductors~\cite{Schnyder_et_al-2008}. Class BDI
is indexed by a topological invariant $\nu \in \mathbb Z$.
(Interactions, which we do not include here, break the classification
down to $\mathbb Z_8$~\cite{Fidkowski-Kitaev-2010}.)

Besides the number of zero eigenvalues, we can also obtain from $h$ in
Eq.~\eqref{eq:h_junc} the number of wires in which the wave function
has support, i.e., the number of non-zero components of the
$(n_++n_-)-$dimensional eigenvectors with eigenvalue zero. The number
of these non-zero components is given by
\begin{align}
  \rho = \max(n_+,n_-)
  \;.
\end{align}

With the two indices $\nu$ and $\rho$ in hand we can justify the
geometry, Fig.~\ref{fig:brickwall}, that we choose for the wire
network. Notice that three (vertical) wires comprise one of the links
of the brickwall lattice in Fig.~\ref{fig:brickwall}, so the lattice
connectivity is three even though five wires meet at each
junction. Given the polarity assignments of the network, we have at
each site of the brickwall lattice $\nu = |3-2| = 1$ and $\rho =
\max(3,2)=3$. These values allow us to satisfy both conditions (i) and
(ii) above: we have a single zero mode in each site ($\nu=1$) and that
mode leaks through three wires, which connect to three sites
($\rho=3$). (Notice that the polarity assignments in the network are
such that each of the three links connecting to a site has at least
one wire with polarity in the majority set for that site, and hence
the wave function of a mode leaks in the direction of all three
neighboring sites.)

We remark that it is not possible to build a nanowire network that
only uses three-wire junctions because one cannot satisfy both
conditions (i) and (ii) simultaneously. For example, $n_+=2$ and
$n_-=1$ yield a single mode at a site, but then the wave function
would only leak to two out of three neighboring sites. In this
example, instead of a system with 2D connectivity, the network would
behave as a set of decoupled 1D systems.

The arguments above are our rationale for proposing the design in
Fig.~\ref{fig:brickwall} as a way to realize the 2D artificial
``Majorana graphene'' platform on which to build logical Majorana zero
modes. In the next sections, we present a detailed analysis of the
wire network using an electronic tight-binding model of the nanowires
and junctions.

\section{Time-reversal symmetric superconducting nanowires and their
  junctions}
\label{sec:wires_and_junctions}

\subsection{Time-reversal symmetric Kitaev chains}
\label{sec:TRS Kitaev}

Majorana modes appear as zero-energy excitations in a spinless
one-dimensional p-wave superconductor, as shown by Kitaev
\cite{kitaevUnpairedMajoranaFermions2001a}. The Hamiltonian model for
a 1D chain based on Kitaev's idea is given by
\begin{eqnarray}
  H_{\rm Kitaev} & = & \sum_{l = 1}^{L - 1} \left[ - t (a_l^{\dagger}
    a_{l + 1} + a^{\dagger}_{l + 1} a_l) + | \Delta | (e^{i \phi} a_l
    a_{l+ 1} \right. \nonumber \\ & & \left. +\ e^{- i \phi} a_{l +
      1}^{\dagger} a_l^{\dagger}) \right] -\mu \sum_{l = 1}^L \left(
  a_l^{\dagger} a_l - \frac{1}{2} \right),
  \label{eq:HKitaev}
\end{eqnarray}
where $a^\dagger_{l},a_{l}$ are spinless fermion creation and
annihilation operators, $|\Delta|$ and $\phi$ are the chain's
superconductor order parameter amplitude and phase, respectively, $t$
is the hopping amplitude between the nearest-neighbor sites, and $\mu$
is the chemical potential. For this Hamiltonian, provided that
$|\mu|<2t$, it can be shown that the two Majorana zero modes are
localized at both endpoints of the chain within a characteristic
length
\begin{align}
\ell_0 = \max(\ell_0^+,\ell_0^-)
  \;,
  \label{eq:locallength}
\end{align}
where
\begin{align}
\ell_0^{\pm} = \left| \ln \frac{|-\mu \pm
       \sqrt{\mu^{2}-4t^2+4|\Delta|^2}|}{2(t + \Delta)} \right|^{-1}.
\end{align}
Thus, for a finite chain there is always some residual interaction
between the two zero modes at the chain endpoints.

\begin{figure}[h]
  \raisebox{0.0\height}{\includegraphics[width=7.9cm,height=1.19cm]{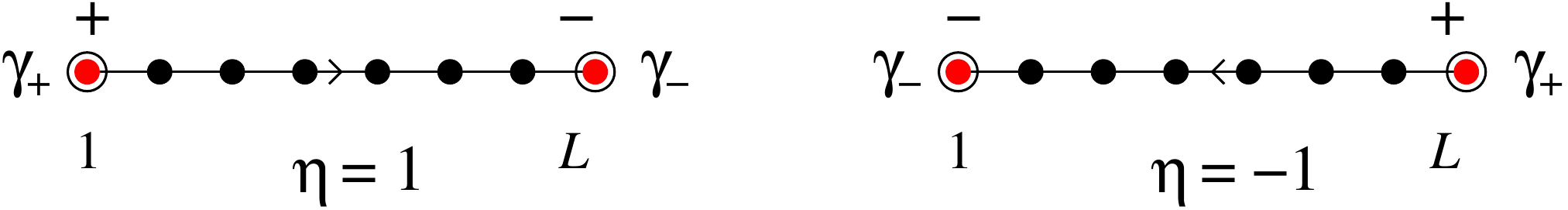}}
  \caption{Schematic illustration of the concept of polarity for
    time-reversal symmetric Kitaev chains. $\eta = \pm 1$ correspond
    to whether the Majorana zero modes at the chain endpoints (here
    indicated by $\gamma_{\pm}$)are even/odd or odd/even under
    time-reversal symmetry.}
    \label{fig:polarity}
\end{figure}

Under the TRS symmetry operation, the spinless fermionic operators
transform as
\begin{eqnarray}
  {\cal T}\, a_l\, {\cal T}^{- 1} & = & a_l \\ {\cal T}\,
  a_l^{\dagger}\, {\cal T}^{- 1} & = & a_l^{\dagger},
\end{eqnarray}
while a scalar $z$ transforms as
\begin{equation}
  {\cal T}\, z\, {\cal T}^{- 1} = z^{\ast} .
\end{equation}
Therefore, applying the TRS operation on the chain Hamiltonian we
obtain
\begin{eqnarray}
  {\cal T}\, H_{\rm Kitaev}\, {\cal T}^{- 1} & = & \sum_{l = 1}^{L -
    1} \left[ - t (a_l^{\dagger} a_{l + 1} + a^{\dagger}_{l + 1} a_l)
    \right. \nonumber \\ & & \left. +\ | \Delta | (e^{- i\phi} a_l
    a_{l + 1} + e^{i \phi} a_{l + 1}^{\dagger} a_l^{\dagger}) \right]
  \nonumber \\ & & -\ \mu \sum_{l = 1}^L \left( a_l^{\dagger} a_l -
  \frac{1}{2} \right).
\end{eqnarray}
We can easily verify that in order for the Hamiltonian to be
time-reversal symmetric we must have $\phi = \pi n$, where $n = 0,\pm
1, \pm 2, \ldots .$ In this case, we can write $\Delta = \pm | \Delta
|$ (positive or negative). However, since $a_l a_{l + 1} = - a_{l + 1}
a_l$, we can turn a ``negative'' sign in $\Delta$ into a positive one
by running the index $l$ from $L$ to 1 instead of 1 to $L$. Therefore,
the orientation of the hopping in the superconductor term and the sign
of $\Delta$ are related. We can take this into account by classifying
time-reversal symmetric Kitaev chain into two classes: ``right'' and
``left''. Thus, in general, for time-reversal symmetric chains we have
\begin{eqnarray}
  H_{\rm Kitaev} & = & \sum_{l = 1}^{L - 1} \left[ -t (a_l^{\dagger}
    a_{l + 1} + a^{\dagger}_{l + 1} a_l) \nonumber \right. \\ & &
    \left. +\ \eta | \Delta | (a_l a_{l + 1} + a_{l + 1}^{\dagger}
    a_l^{\dagger}) \right] \nonumber \\ & & -\ \mu \sum_{l = 1}^L
  \left( a_l^{\dagger} a_l - \frac{1}{2} \right),
    \label{eq:HKitaevTRS}
\end{eqnarray}
where $\eta = \pm 1$. In fact, we can introduce the concept of chain
``polarity'', see Fig. \ref{fig:polarity}, where $\pm$ signs are
associated to the endpoints of the chain (i.e, site coordinates $l =
1$ or $l = L$), as well as an arrow, depending on the sign of
$\eta$. As we showed in Sec. \ref{sec:indices}, the polarity of the
chain is connected to how the Majorana zero modes at the chain
endpoints ends transform under time reversal.

Although Kitaev's model describes a platform for the realization of
Majorana zero modes in one spatial dimension, electrons have spin and
most superconductors occurring in nature have s-wave pairing. More
realistic proposals~\cite{oregHelicalLiquidsMajorana2010a,
  lutchynMajoranaFermionsTopological2010a,
  aliceaNonAbelianStatisticsTopological2011a} obtain a topological
phase and Majorana boundary modes by including three ingredients: (i)
proximitized s-wave superconductivity, (ii) Rashba spin-orbit
coupling, and (iii) a Zeeman field.

The continuum Hamiltonian for a single-channel nanowire with electron
effective mass $m^\ast$, Rashba spin-orbit coupling $\lambda$ and
applied Zeeman field $B_{z}$, proximitized with an s-wave
superconductor with a pairing amplitude $\Delta_s$, with a chemicalk
potential $\mu_w$ can be written as
\begin{eqnarray}
  H_{\rm wire} & = &\frac{1}{2} \int dx\, \Psi (x)^{\dagger} \left[
    \left( -\frac{\hbar^2 \partial _x^2 }{2 m^{\star}} - i \lambda
    \partial_x \sigma_y - \mu_w \right) \tau_z \right. \nonumber \\ &&
    \left. +\ \frac{g \mu_B | B_{z} |}{2} \sigma_z + \Delta_s \tau_x
    \right] \Psi (x)
  \label{eq:Hcontinuum}
\end{eqnarray}
in the Bogoliubov-De Gennes (BdG) formulation. Here we use the Nambu
spinor formulation where
\begin{equation}
    \Psi(x)^T = (\psi_{\uparrow}(x), \psi_{\downarrow}(x),
    \psi^{\dagger}_{\downarrow}(x),
    -\psi^{\dagger}_{\uparrow}(x))
\end{equation}
and $\psi_\sigma^\dagger(x)$ and $\psi_\sigma(x)$ are fermionic
creation and annihilation field operators. As discussed in
Ref.~\onlinecite{aliceaNonAbelianStatisticsTopological2011a}, it can
be shown that within the limits $|\Delta| < g \mu_B | B_{z} |/2 $ and
$g \mu_B | B_{z} | \gg m^{\star} \lambda^2$ the Hamiltonian in
Eq.~\eqref{eq:Hcontinuum} can be projected to a single-band effective
low-energy Hamiltonian that matches Kitaev's model in
Eq.~\eqref{eq:HKitaev}. Upon diagonalization, it can be shown that the
nanowire Hamiltonian in Eq.~\eqref{eq:Hcontinuum} can be driven to a
topological phase when
\begin{equation}
  \frac{g \mu_B | B_{z} |}{2} > \sqrt{\Delta_s^2 + \mu_w^2}.
  \label{eq:critical_B}
\end{equation}

Experimentally, the chirality of the nanowire can be controlled by
strain, which defines the sign of the Rashba spin-orbit coupling in
Eq.~\eqref{eq:Hcontinuum}. The direction of the effective electric
field due to the strain enters into the spin-orbit Hamiltonian as
follows
\begin{equation}
H_{\rm Rashba} = \frac{g\mu_{B}} {2mc} \left( {\bf \sigma} \times
\mathbf{p} \right) \cdot {\bf E}_{\rm eff}.
\label{eq:Rashab_spin_orbit}
\end{equation}
The interplay between the spin-orbit coupling and the s-wave
superconductivity makes the sign of the effective p-wave order
parameter directly depends on the relative direction of the
strain-induced electric field with respect to a fixed crystal
direction \cite{aliceaNonAbelianStatisticsTopological2011a}, namely,
\begin{equation}
\Delta \cong \frac{\lambda \cdot \Delta_{s}} {g\mu_{B} B_{z}} \sim
\text{sgn} \left( {\bf E}_{\rm eff} \cdot \hat{z} \right).
\end{equation}

Given the strong connection between the physics encoded in the
realistic model represented by Eq.~(\ref{eq:Hcontinuum}) and the
Kitaev Hamiltonian in the topological regime, in this paper we adopted
the latter for our modeling and analysis of nanowires. This assumption
simplifies the calculations without affecting the generality of our
conclusions. In Sec.~\ref{sec:parameters}, we return to the continuum
model to connect our results to the experimental parameter space.

\subsection{Electronic tight-binding model of nanowires and junctions}
\label{sec:tight-binding}

We start by investigating junctions of time-reversal symmetric Kitaev
nanowires. We only consider junctions in the all-connected
configuration, where all nanowire endpoints are coupled to each other.

We employ an electronic, single-band electronic tight-binding model
based on the Hamiltonian of Eq. (\ref{eq:HKitaev}) to model nanowires
and their junctions. We adopt a BdG representation for the fermion
operators
\begin{equation}
\psi_{l,\alpha} = \left( \begin{array}{c} a_{l,\alpha}
  \\ -a_{l,\alpha}^\dagger \end{array} \right), \quad
\psi^\dagger_{l,\alpha} = \left( \begin{array}{cc}
  a_{l,\alpha}^\dagger & -a_{l,\alpha} \end{array} \right),
\end{equation}
where $l=1,\ldots,L$ is the site coordinate and $\alpha=1,\ldots,n$ is
the nanowire index. We number the sites by starting from the junction
end of the nanowire. The total Hamiltonian of a $n$-nanowire junction
system is written as
\begin{equation}
  H_{\rm total} = \sum_{\alpha=1}^n H_{\alpha} + H_{\rm
    junction},
\label{eq:Htotal}
\end{equation}
where
\begin{eqnarray}
H_{\alpha} & = & \sum^{L-1}_{l = 1} \psi^\dagger_{l,\alpha}
\left( - t\, \tau_z + i \eta_\alpha | \Delta | \tau_y \right)
\psi_{l+1,\alpha} \nonumber \\ & & -\frac{\mu}{2} \sum^{L}_{l=1}
\psi^\dagger_{l,\alpha}\, \tau_z\, \psi_{l,\alpha}
\label{eq:H_wire}
\end{eqnarray}
describes the $\alpha$-th wire and
\begin{eqnarray}
  H_{\rm junction} &=& -\frac{1}{2} \sum_{\alpha\neq\beta}
  \Gamma_{\alpha\beta} \psi_{1,\alpha}^\dagger\, \tau_z\,
  \psi_{1,\beta}
  \label{eq:H_Y}
\end{eqnarray}
describes the couplings at the junction, where $\Gamma_{\alpha\beta}$
is the pairwise hopping amplitude between the endpoints of the
$\alpha$-th and $\beta$-th nanowires. For the simulations discussed in
this section, we adopt $\mu = 0.5 t$, $L=20$ and $\Delta = 0.5 t$,
which sets the nanowires in the topological regime and
exponentially localize the Majorana zero modes at the nanowire ends,
with $\ell_0 \approx 1.82$. We have performed all the numerical
simulations in this paper by implementing the tight-binding
Hamiltonian of Eq. (\ref{eq:Htotal}) in Kwant
\cite{grothKwantSoftwarePackage2014a}.

\begin{figure*}
  \raisebox{0.0\height}{\includegraphics[width=0.8\textwidth]{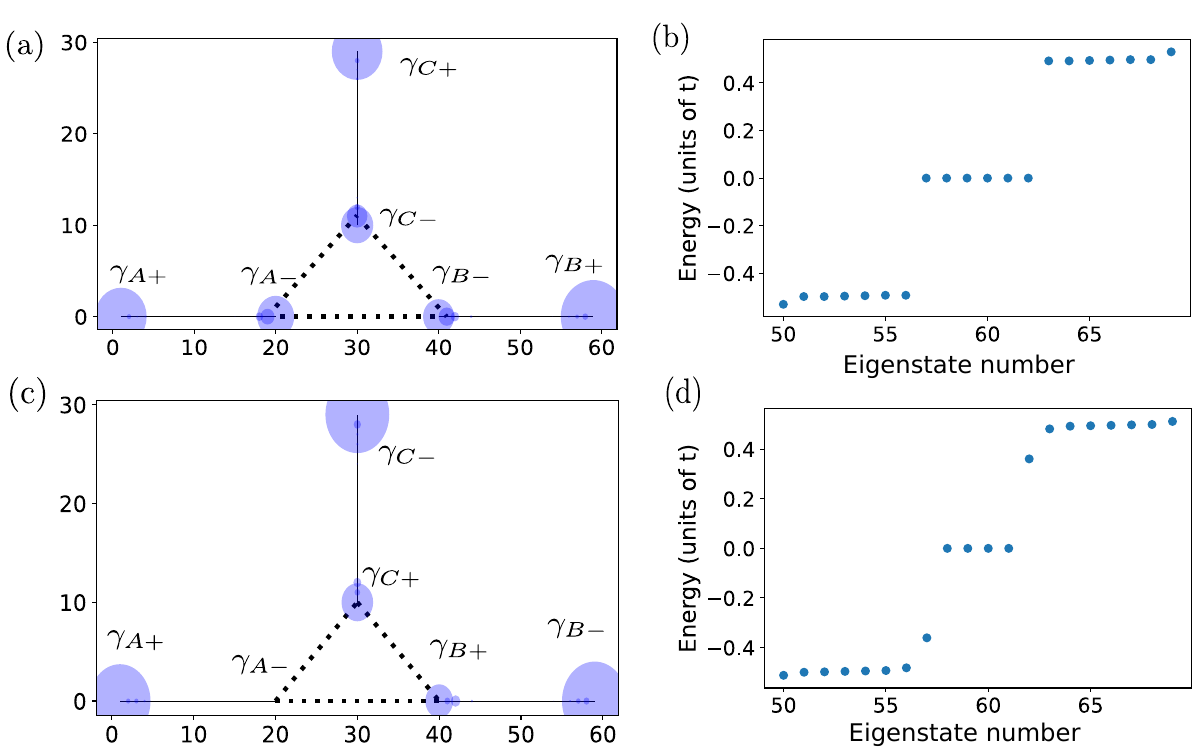}}
  \caption{Numerical results for a three-wire junction with two
    different choices of polarity. Solid lines indicate the location
    of the nanowires (site coordinates are shown), while dotted lines
    connect neighboring nanowire endpoints. To facilitate
    visualization, neighboring endpoints are set farther apart than
    one chain lattice constant unit. In panels (a) and (b) the
    electronic local density of states (LDOS) at zero energy and the
    energy eigenstates are plotted, respectively, for the $(- - -)$
    case, while in panels (c) and (d) the same quantities are plotted
    for the $(+ + +)$ case. The wires are identified by the subscripts
    $A,B,C$. In (a), it is noticeable that the zero-energy wave
    function has amplitude in the three junction sites and the total
    number of zero modes is six, with three sitting at the junction
    and three other modes sitting at the outer ends of the wires. In
    (c), the zero-energy wave function has amplitude in the majority
    polarity sites of the junction but there is only zero mode at the
    junction. The other three modes sit at the outer ends of the
    nanowires.}
  \label{fig:Y-junction}
\end{figure*}

\subsection{Majorana junctions of 3 wires}
\label{sec:junctions}

We note that for two Majorana zero modes on different time-reversal
symmetric nanowires to hybridize and to combine into a finite-energy
fermion, they must to be of different polarity. This restricts how the
zero modes can be distributed among the junction nanowires. In the
case of a three-wire junction, two possible cases exist. When all the
Majorana zero modes in the junction are of the same polarity, none of
the zero modes couple and all six zero modes on the ends of three
nanowires survive. However, when one of the nanowires has a different
polarity than the other two nanowires, for example, in a $(+ + -)$ or
$( - + -)$ configuration, there is only one zero mode at the junction
and the wave function amplitude for that zero mode is shared between
the majority polarization sites.

This analysis in terms of Majorana operators is corroborated by a
numerical simulation of the underlying electronic system. The results
are presented in Fig.~\ref{fig:Y-junction}, where the electronic local
density of states (LDOS) at zero energy for the $(- - -)$ and $(- + +
)$ junction configurations are shown when $\Gamma_{\alpha\beta} =
(1-\delta_{\alpha,\beta})\Gamma$ with $\Gamma=t$. It is clear that for
the $( - - - )$ configuration a total of six Majorana zero modes are
present, including three at the junction. For the $(- + + )$
configuration, there is a single zero mode at the junction and it is
shared only by the nanowires with majority polarity.

\begin{figure}
  \raisebox{0.0\height}{\includegraphics[width=6cm]{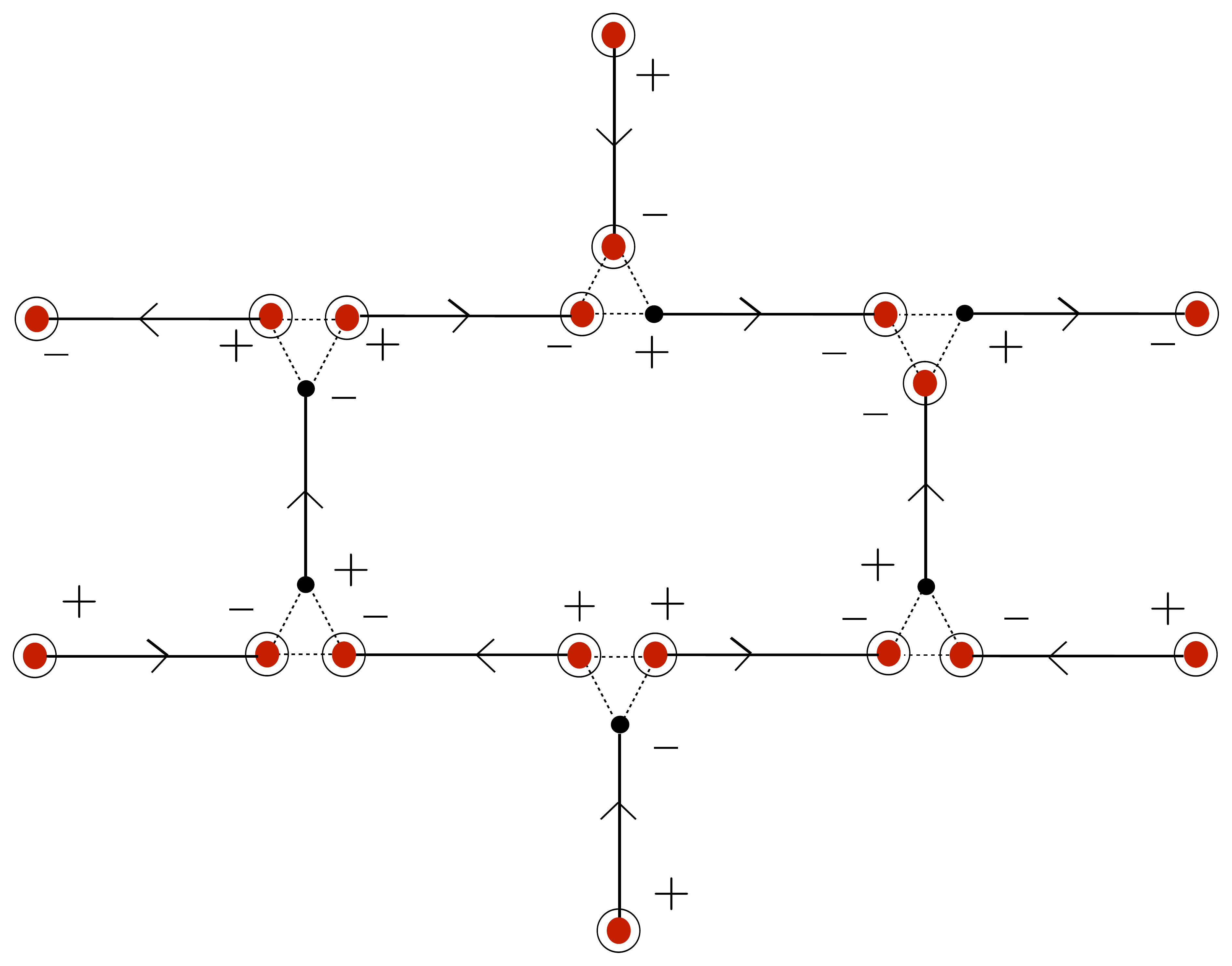}}
  \caption{Schematic illustration of a brickwall network of nanowires
    with well-defined polarizations and three-wire junctions of the
    $(- + +)$ and $(+ + -)$ types. The presence of a nonzero
    Majorana zero mode amplitude at a nanowire end is indicated by a
    circular line around a solid circle. Nanowire polarities are
    indicated by arrows and $\pm$ signs.}
  \label{fig:honeycomb-cell}
\end{figure}

The wave function distribution in the majority-polarized nanowires
makes it impossible to satisfy simultaneously the two necessary
conditions for the realization of a single-band Majorana network. This
illustrated in Fig.~\ref{fig:honeycomb-cell} where a brickwall
network out of the nanowires with only one Majorana zero mode at each
junction is shown. The Majorana zero mode located on the majority
nanowires of a junction is disconnected from the zero mode on the
junction across the minority nanowire. As a result, when hybridization
within the nanowires is turned on, the network breaks up into an array
of disconnected chains with no inter-chain coupling.

\subsection{Majorana junctions of five nanowires}
\label{sec:junctions_more}

\begin{figure*}
  \raisebox{0.0\height}{\includegraphics[width=0.8\textwidth]{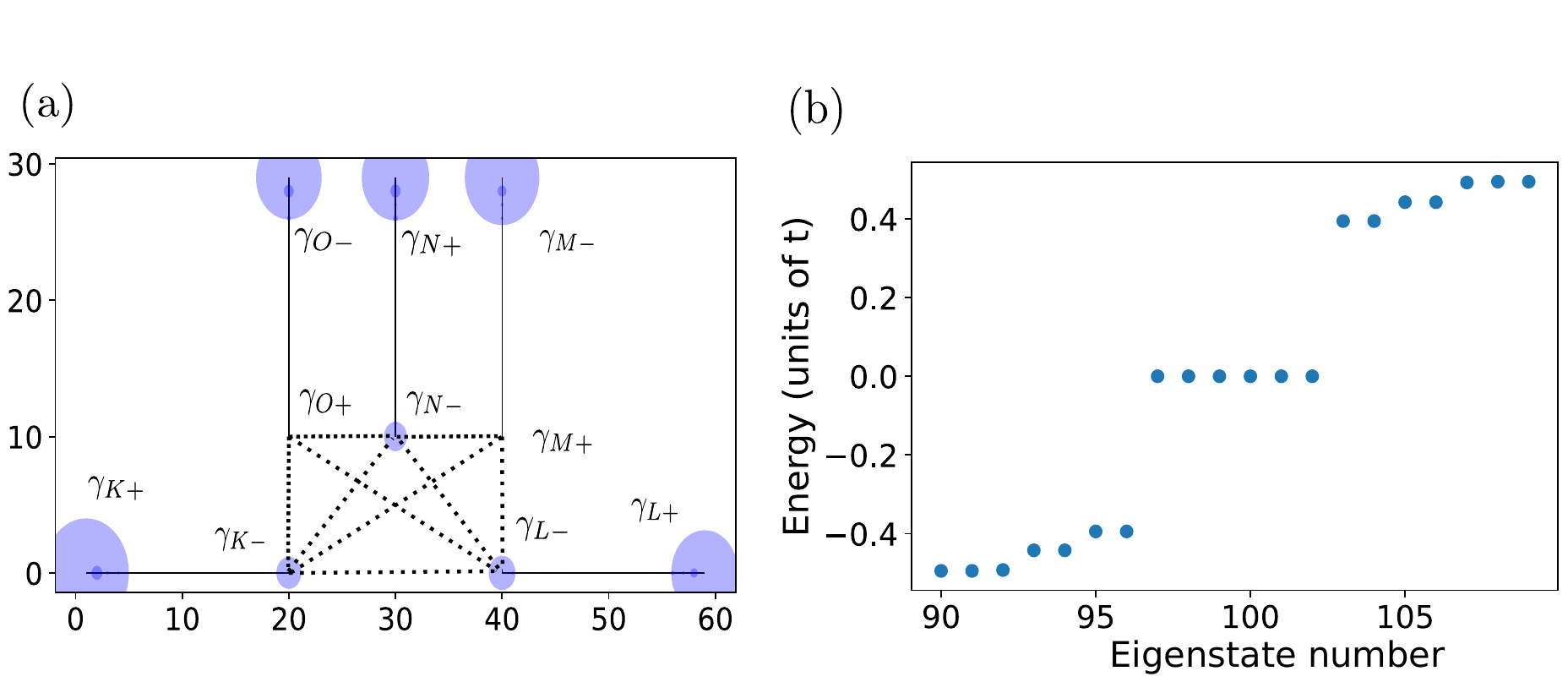}}
  \caption{Numerical results for a five-wire junction consisting of
    two positive and three negative polarity nanowires coupled
    according to Eq. (\ref{eq:5junction}). The same conventions as in
    Fig. \ref{fig:Y-junction} are followed here. In panel $(a)$, the
    electronic local density of states (LDOS) at zero energy is
    plotted while in panel $(b)$ the relevant energy eigenstates are
    plotted. The wires are identified by the subscripts
    $K,L,M,N,O$. It is clear from the electronic LDOS that the zero
    energy wave function in the junction has nonzero amplitude only in
    the majority polarity sites. There is no impediment for the zero
    mode located at a junction can hybridize with the zero modes in
    neighboring junctions.}
  \label{fig:5-wire-junction}
\end{figure*}

We now consider the case of five nanowires and when $n_+ = 2$ and $n_-
= 3$. We simulate such a junction similarly to the three-nanowire
junction using the following choice for the $\Gamma_{\alpha \beta}$
coupling parameters:
\begin{equation}
  \Gamma_{\alpha\beta} = \frac{\Gamma}{\sqrt{6}} \times
  \left( \begin{array}{ccccc} 0 & 0 & 1 & 1 & -2 \\ 0 & 0 & \sqrt{3}
    & -\sqrt{3} & 0  \\ 1 & \sqrt{3} & 0 & 0 & 0 \\ 1 & -\sqrt{3} &
    0 & 0 & 0 \\ -2 & 0 & 0 & 0 & 0 \end{array} \right).
  \label{eq:5junction}
\end{equation}
%
%
This choice of coupling matrix elements lifts all by one mode from
zero energy. The remaining zero mode has equal amplitude among the
three majority-polarity nanowires. The results are shown in
Fig.~\ref{fig:5-wire-junction}. In this case, the single Majorana zero
mode wave function is distributed among the negative endpoints of the
majority polarization nanowires in the junction, fully connecting the
three links of the network that emanate from the junction and
condition (i) of Sec. \ref{sec:indices} is satisfied. Moreover, all
junction zero modes can be connected to zero modes located at the
outer endpoints of the junction nanowires, satisfying condition
(ii). This configuration is therefore chosen for the simulation of a
``Majorana graphene'' network, which we discuss in the following
section. While others choices of coupling matrix elements are
possible, as long as the diagonal blocks in Eq. (\ref{eq:5junction})
are zero, only one zero mode exists at the junction. Small deviations
from the coupling matrix elements in Eq.~(\ref{eq:5junction}) are
possible: due to TRS, Dirac cones in the dispersion relation survive
as long as the matrix elements satisfy the triangle rule, which says
that the magnitude of each of the tunnel couplings is smaller than the
sum of the magnitudes of the other two couplings
\cite{Bernevig_2013}. Moreover, Dirac cones are also robust to
inhomogeneities in $\Gamma$ from one junction to another, as there is
no equivalent to on-site disorder in a Majorana ``graphene'' system.

\section{Majorana Network}
\label{sec:network}

In Sec. \ref{sec:junctions_more}, we established that, using a
five-wire junction, it is possible to (i) have a single Majorana
(quasi-)zero mode per site of the brickwall lattice, and (ii)
hybridize this mode with those on the three neighboring sites. These
results justify using the brickwall structure in
Fig.~\ref{fig:brickwall} to realize the ``Majorana graphene'' network.
Here we proceed to construct the logical Majorana zero mode using this
structure, modeling every nanowire in the network by the Kitaev
Hamiltonian in the BdG formulation, as in Eq. (\ref{eq:H_wire}). (In
the next section we discuss the connection to a more experimentally
realistic model.)

\subsection{``Majorana graphene'' network}

The characteristic length of the zero modes in the Kitaev nanowires
depends on the chemical potential, as indicated in
Eq. (\ref{eq:locallength}). It is thus possible to increase the
overlap between the zero modes in neighboring sites of the network,
i.e., effective hopping of the Majorana zero modes in the brickwall
lattice, by controlling a gate voltage $V_g$ in every nanowire.

\begin{figure*}
  \raisebox{0.0\height}{\includegraphics[width=0.9\textwidth]{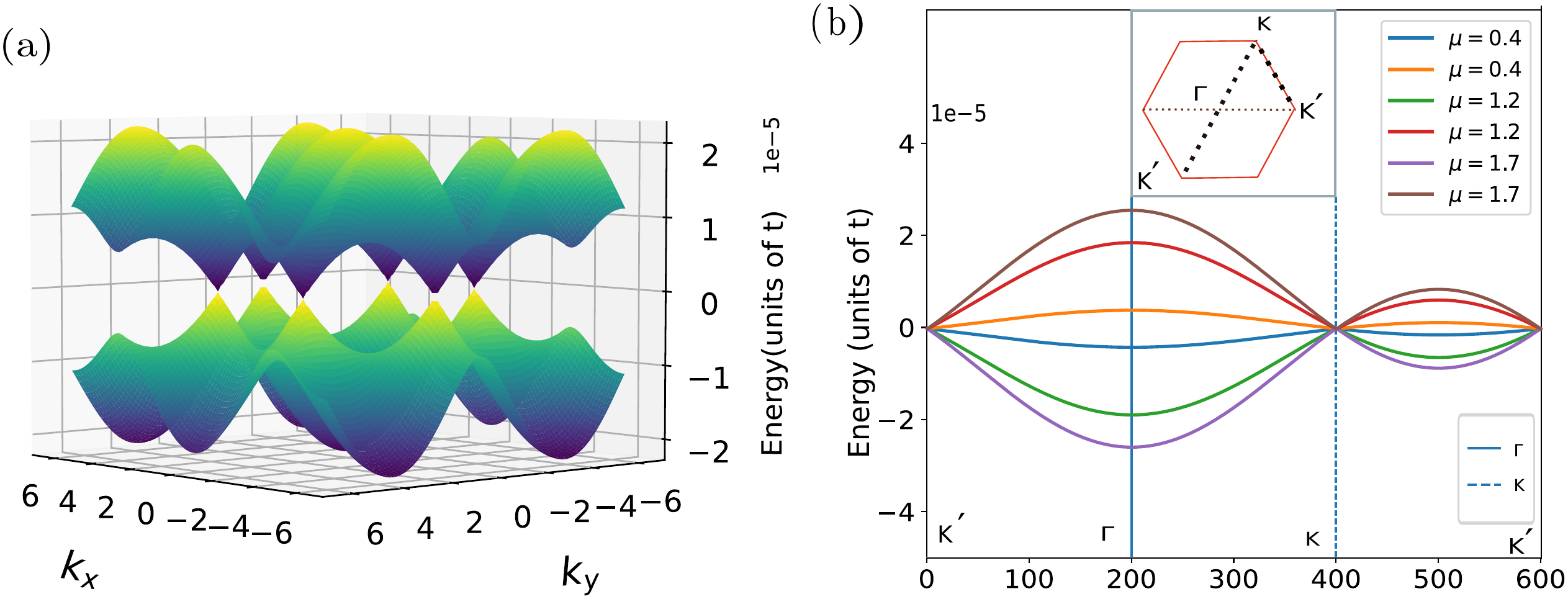}}
  \caption{(a) Electronic energy bands near zero energy for an
    infinite network of five-wire junctions consisting of two positive
    and three negative polarity nanowires. The nanowire parameters are
    $L=20$, $\Delta = 0.5t$, and $\mu=0.5t$. The junction coupling
    parameters are chosen according to Eq. (\ref{eq:5junction}) with
    $\Gamma=t$. The brickwall lattice is reshaped as a honeycomb
    lattice in order to create a triangular reciprocal lattice unit
    cell to facilitate visualization of the bands. $(b)$ Energy bands
    along the reciprocal space dashed line path are shown in the
    inset. Bands are various chemical potential values are shown to
    illustrate their impact on the Fermi velocity at the $K_\pm$
    points.}
  \label{fig:7-dirac}
\end{figure*}

We consider first the case with uniform hopping matrix elements (i.e.,
uniform gate voltages) across the entire network. In this case, the
Majorana system on the brickwall lattice contains features similar to
those of graphene, such as a Dirac-type dispersion. To illustrate this
point, we computed the electronic energy bands of an infinite
honeycomb network of 5-nanowire junctions. In Fig.~\ref{fig:7-dirac}a
we show the energy bands close to zero energy, where the six pairs of
Dirac cones at the $K_\pm$ points are clearly visible. The nanowire
parameters are $L=20$, $\Delta = 0.5t$, and $\mu = 0.5t$; the junction
couplings follow Eq. (\ref{eq:5junction}), with $\Gamma = t$. 
  In Fig.~\ref{fig:7-dirac}b, we show that the Fermi
velocity for the Dirac dispersion depends on the nanowire chemical
potential, which in turn controls the effective hopping amplitude
between Majorana zero modes located at the opposite ends of the
nanowire.

\begin{figure}[h]
  \raisebox{0.0\height}{\includegraphics[width=8cm]{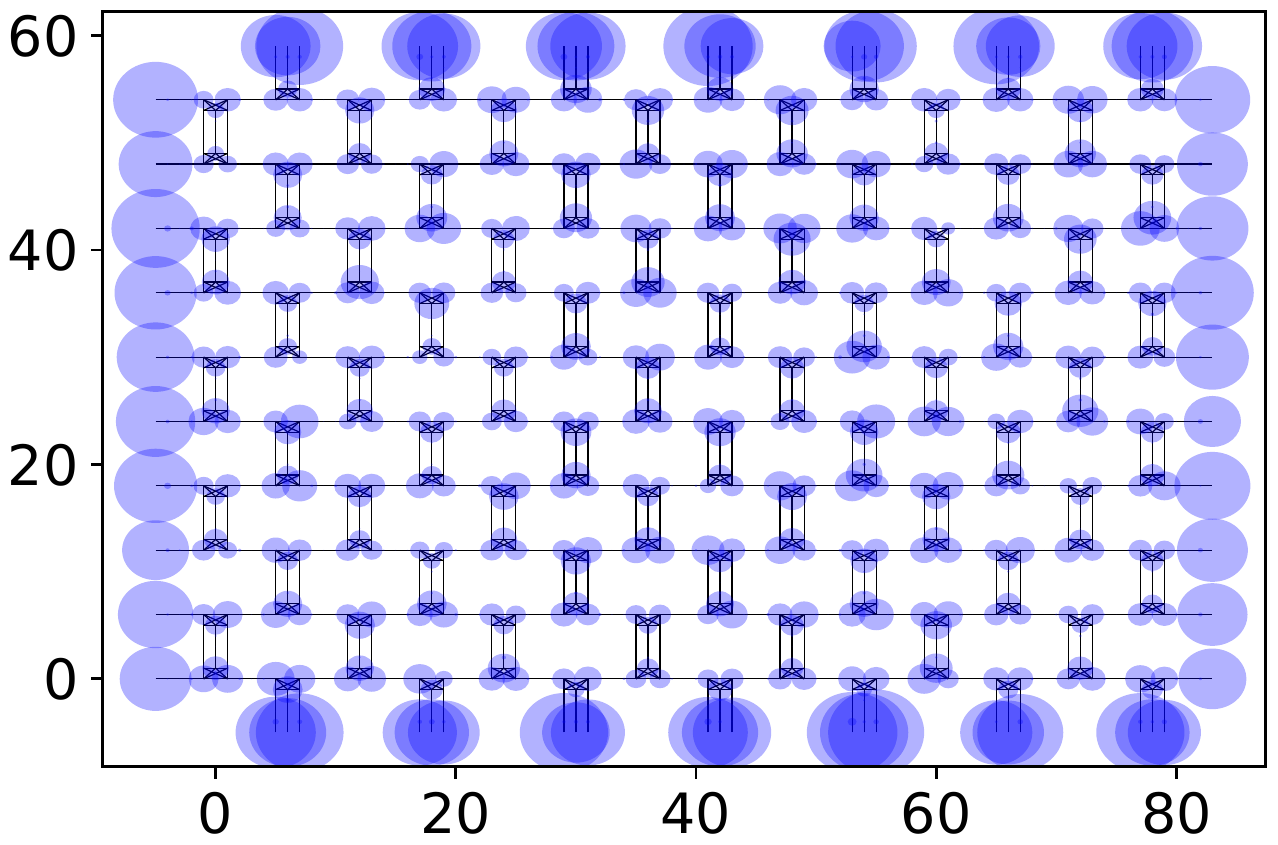}}
  \caption{Electronic LDOS for a $7\times 10$ brickwall lattice of
    Majorana nanowires with open boundary conditions and a single zero
    mode at each junction. Horizontal links in the lattice consist of
    a single nanowire while vertical links contain three wires. Every
    nanowire is described by a Kitaev chain Hamiltonian with $L=5$
    sites, $\mu=0.4t$, and $\Delta=0.8t$. Junction couplings follow
    Eq. (\ref{eq:5junction}) with $\Gamma = t$.}
  \label{fig:majorana-lattice}
\end{figure}

\begin{figure}[h]
  \raisebox{0.0\height}{\includegraphics[width=8cm]{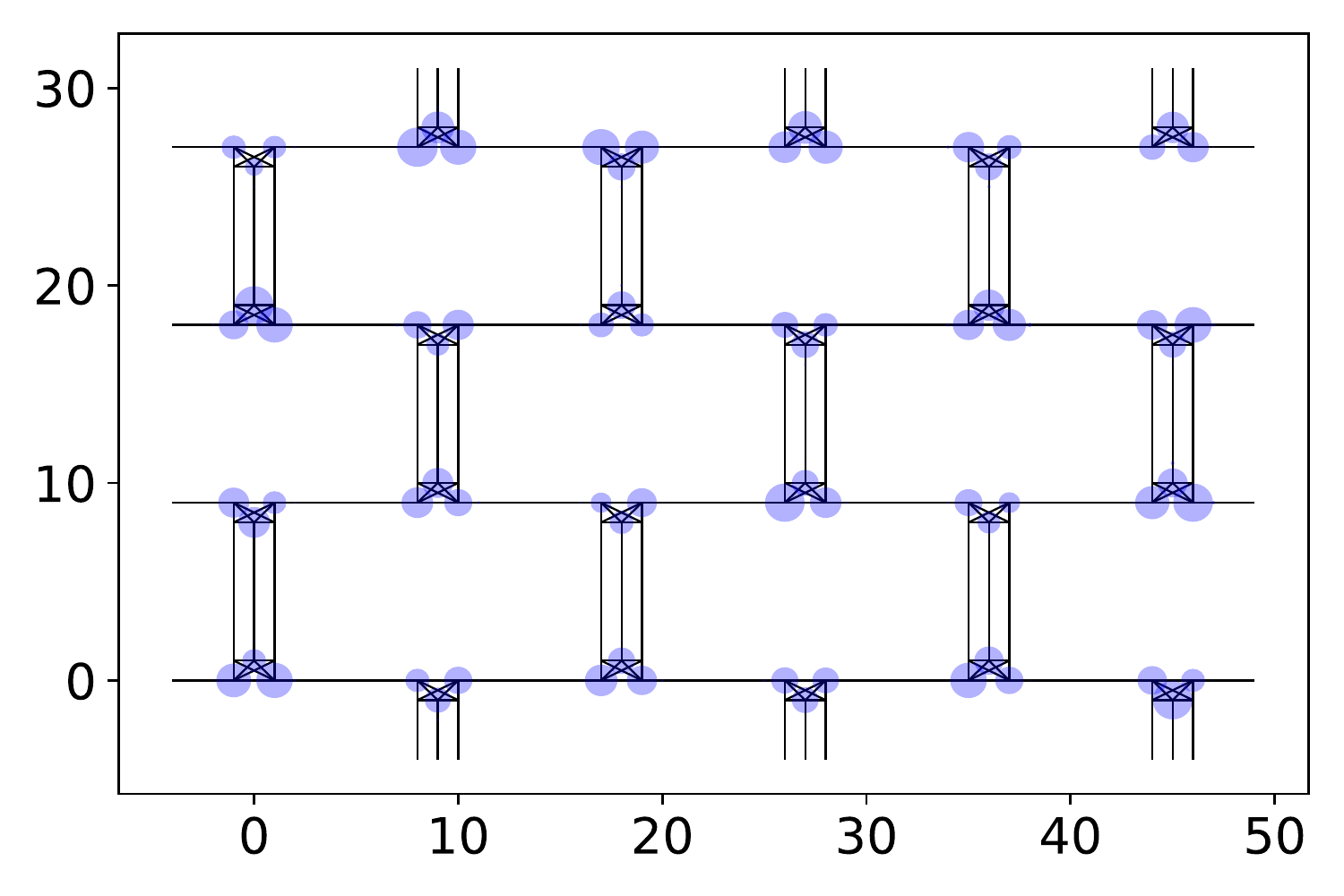}}
  \caption{Electronic LDOS for a $3\times 4$ brickwall lattice of
    Majorana nanowires similar to that of
    Fig. \ref{fig:majorana-lattice} but with periodic boundary
    conditions and $L=8$. Notice the absence of zero modes at the
    boundaries. The smaller lattice has been used to show that the
    boundary Majorana-zero modes disappear when using periodic
    boundary conditions.}
  \label{fig:majorana-lattice-periodic}
\end{figure}

Returning to the finite-size network of Fig.~\ref{fig:brickwall} with
open boundaries, in Fig.~\ref{fig:majorana-lattice} we show its
electronic LDOS at zero energy for the network when $\mu = 0.4t$,
$\Delta=0.8t$, and $\Gamma = t$ , employing the junction
coupling matrix of Eq. (\ref{eq:5junction}). The LDOS shows zero
energy modes at the boundary and in the bulk of the system. These zero
modes correspond to the states at the Dirac nodes (i.e., the apexes of
the cones in the energy bands of Fig. \ref{fig:7-dirac}). Due to the
open boundary conditions, zero modes appear at the boundary
sites. They can be removed by switching to periodic boundary
conditions, as we show in Fig. \ref{fig:majorana-lattice-periodic}.

\subsection{Kekul\'{e} modulation in the brickwall lattice}
\label{sec:kekule}

\begin{figure}[h]
\raisebox{0.0\height}{\includegraphics[width=6cm]{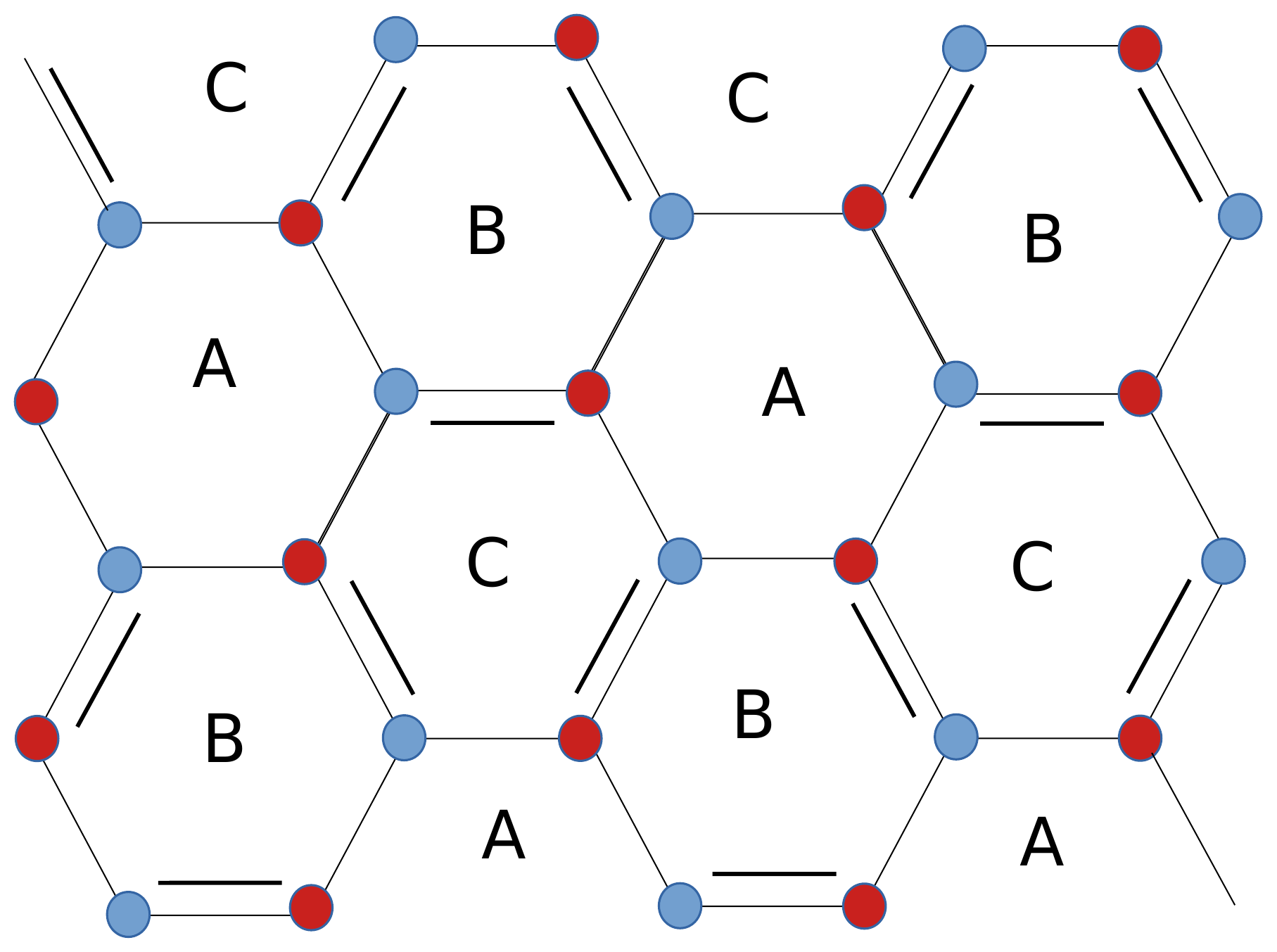}}
\caption{The Kekul\'{e} dimerization in a honeycomb lattice. The
  single (double) links correspond to weak (strong) bond
  amplitudes. The red and blue dots represent the two sublattices of
  the honeycomb lattice. As a result of the dimerization, three kinds
  of plaquettes are created which are labeled by A, B and C.}
  \label{fig:Kekule_honeycomb}
\end{figure}

Introducing a Kekul\'{e} dimerization pattern in a graphene lattice
opens up a gaps in the Dirac spectrum
\cite{chamonSolitonsCarbonNanotubes2000,
  yangHierarchicalMajoranasProgrammable2019a}. The Kekul\'{e}
modulation can be realized by imposing the following perturbation to
the local chemical potential (via gate voltages):
\begin{equation}
  \mu = \mu_0 + \delta \mu_{\mathbf{r}, \alpha},
  \label{eq_kekule1}
\end{equation}
where
\begin{equation}
  \delta \mu_{\mathbf{r}, \alpha}^{\rm Kekule} = \mu_K \cos \left(
  \varphi_{\mathbf{r},\alpha} \right).
  \label{eq_kekule2}
\end{equation}
and
\begin{equation}
\varphi_{\mathbf{r},\alpha} = \mathbf{K_+} \cdot \mathbf{s_{\alpha}} +
(\mathbf{K_+} - \mathbf{K_-}) \cdot \mathbf{r}.
\end{equation}
To implement this modulation, we return momentarily to the equivalent
honeycomb lattice and its coordinate system. The position vector
$\mathbf{r}$ has a fixed (arbitrary) origin and points to the sites of
one of the triangular sublattices. The three vectors $s_{\alpha}
(\alpha = x, y, z)$ connect sites of that sublattice to their nearest
neighbors on the other sublattice. ${\bf K}_+ = - {\bf K}_{ -}$ are
the distinct vectors connecting the $\Gamma$ to the $K$ points in the
reduced Brillouin zone in reciprocal space for the honeycomb
lattice. In Fig. \ref{fig:Kekule_honeycomb} we show the pattern
induced by the Kekul\'{e} distortion.

\begin{figure}
  \raisebox{0.0\height}{\includegraphics[width=8cm]{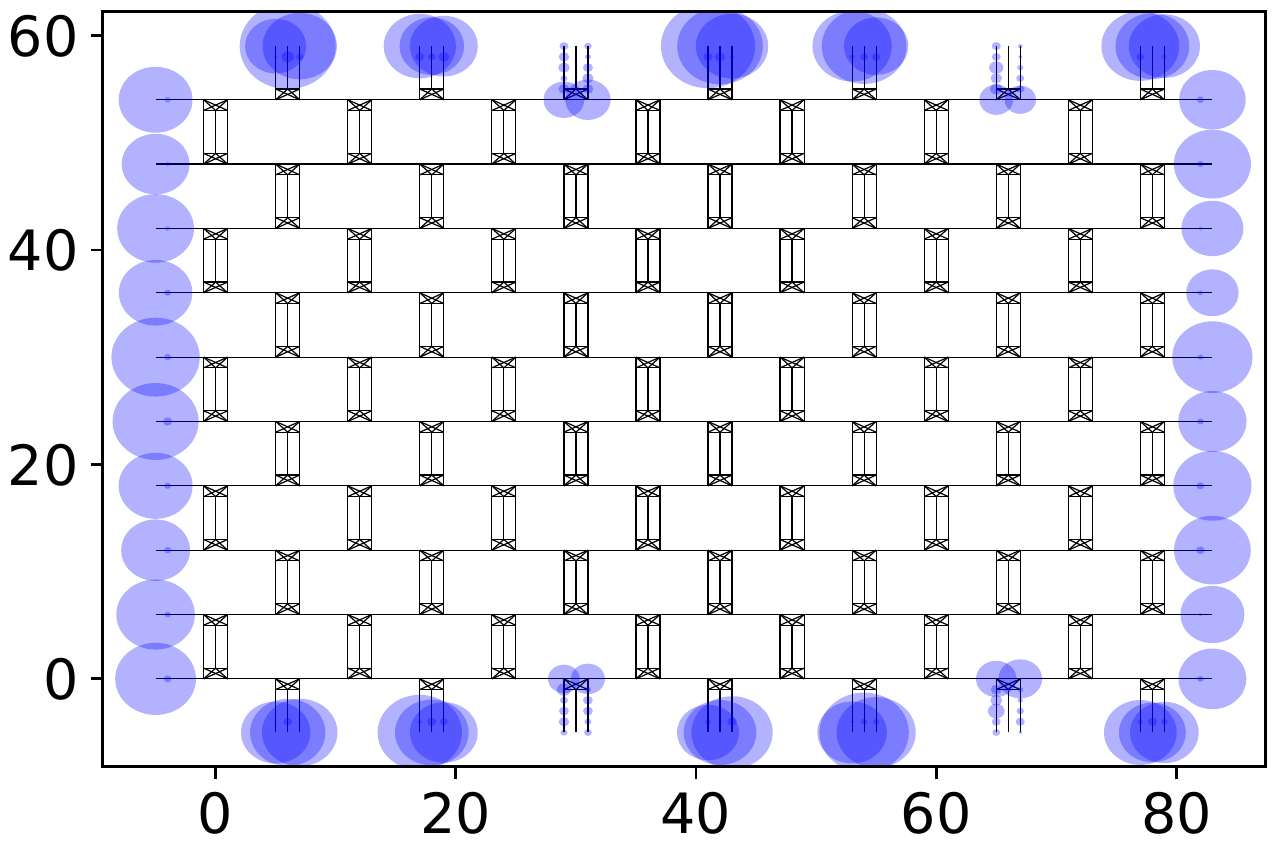}}
  \caption{Electronic LDOS of a brickwall network of Majorana
    nanowires with a Kekul\'{e} distortion on the chemical
    potentials. Chain and junction parameters are the same as in
    Fig. \ref{fig:majorana-lattice}, with the baseline chemical
    potential $\mu_0=t$ and the added Kekul\'{e} modulation
    $\mu_K=0.98t$.}
  \label{fig:gappedlattice}
\end{figure}

\begin{figure*}
  \raisebox{0.0\height}{\includegraphics[width=0.8\textwidth]{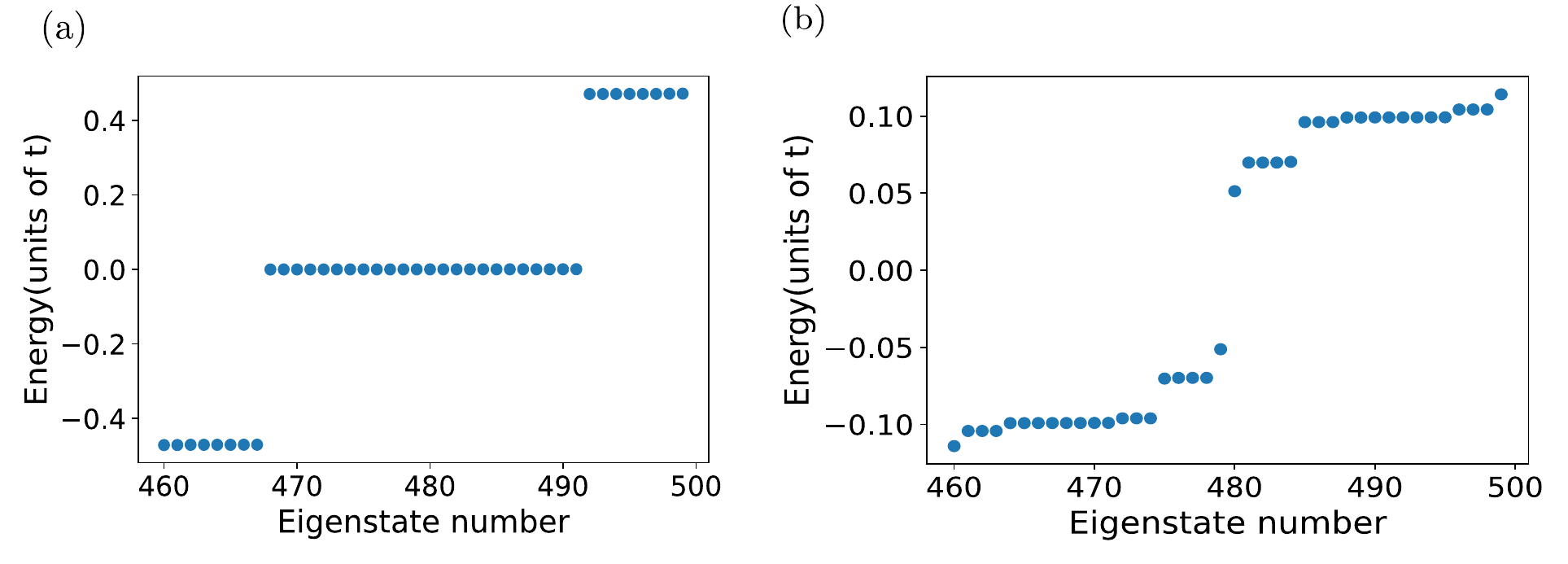}}
  \caption{The energy spectrum of a brickwall lattice with
    quasi-Majorana zero modes at each vertex and periodic boundary
    conditions without (a) and with (b) a Kekul\'{e} modulation for $L
    = 8$ and the same network size as in
    Fig. \ref{fig:majorana-lattice-periodic}. The parameters used in
    panels (a) and (b) are the same as those in
    Fig. \ref{fig:majorana-lattice} and Fig. \ref{fig:gappedlattice},
    respectively.}
  \label{fig:spectrum-gapped}
\end{figure*}

In Fig.~\ref{fig:gappedlattice}, we show the effect resulting from the
Kekul\'{e} dimerization pattern on the electronic LDOS at zero energy
for the wire network in Fig.~\ref{fig:brickwall}. For these
calculations, the junction coupling amplitude $\Gamma = t$, the superconductor order parameter amplitude $\Delta =0.8t$,
and the chemical potential parameters are set as $\mu_0=t$ and
$\mu_K=0.98t$. We choose the maximum value of the chemical potential
$(\mu_{max} = \mu_0 + \mu_K)$ to be very near the boundary of the
topological range, $\mu_{\max} = 1.98t < \mu_c = 2t$ to ensure a
sizable overlap between the Majorana zero modes on both ends of the
same nanowire, with the Majorana characteristic length reaching
$\ell_0 \approx 80$ on those nanowires [see
  Eq. (\ref{eq:locallength})]. Notice that the bulk zero modes are now
absent because of the bulk gap, while the boundary zero modes
remain. The boundary zero modes disappear under periodic boundary
conditions. To illustrate this point, in
Fig.~\ref{fig:spectrum-gapped} we show the energy eigenvalues of a
Majorana zero-mode lattice of 4 layers with
each layer consists of 5 horizontal wires with periodic boundary
conditions in the presence and in the absence of the Kekul\'{e}
modulation. We adopt the same parameters as in
Fig. \ref{fig:majorana-lattice} and Fig. \ref{fig:gappedlattice}.  for
the absence and presence of Kekul\'{e} modulation respectively.

\subsection{Zero modes bound to Kekul\'{e} vortices}
\label{sec:vortices}

The Kekul\'{e} dimerization pattern can support defects in the form of
vortices. As noted in
Ref. \cite{yangHierarchicalMajoranasProgrammable2019a}, a vortex can
be imprinted via an additional modulation of the Kekul\'{e} pattern,
\begin{equation}
  \varphi_{\mathbf{r}, \alpha} = \mathbf{K_+} \cdot
  \mathbf{s_{\alpha}} + (\mathbf{K_+} - \mathbf{K_-}) \cdot \mathbf{r}
  + \varphi_{\mathbf{r}}^{\rm vortex},
  \label{eq:vortex}
\end{equation}
where
\begin{equation}
\varphi_{\mathbf{r}}^{\rm vortex} = \sum^{\nu}_{n = 1} q_n \arg
(\mathbf{r} - \mathbf{R_n}).
\end{equation}
One important advantage of this construction is that the vorticities
$q_n = \pm 1$ ($n = 1, \cdots \nu$) and the positions of the vortices
$\tmmathbf{R_n} $ are also programmable via the applied gate voltage
on each wire. The Kekul\'{e} vortices bind zero energy modes at their
location -- these are the {\it logical} Majorana zero mode. These
logical Majoranas can be moved by applying gate voltages that
correspond to changing the value of $\mathbf{R_n}$ in
Eq.~\eqref{eq:vortex}.

\begin{figure*}
  \raisebox{0.0\height}{\includegraphics[width=0.8\textwidth]{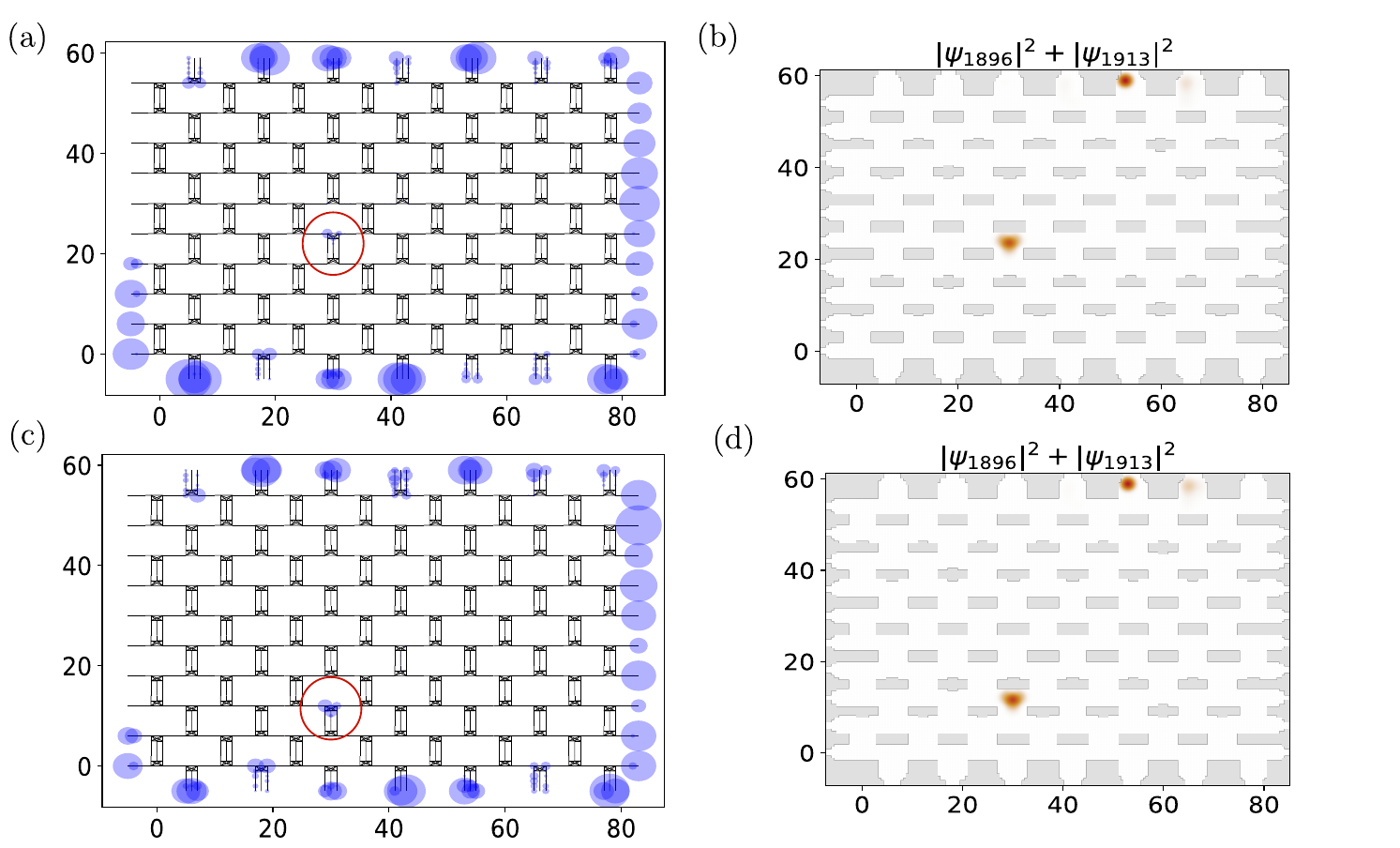}}
  \caption{Vortex in a Majorana network. The electronic LDOS at zero
    energy is plotted in panels (a) and (c) for the two different
    vortex positions indicated by red circles. The intensity of the
    eigenstates located within the red circles is plotted in panels
    (b) and (d). Notice that the eigenstate are the same for both
    vortex positions. The Majorana zero modes in the bulk are bound to
    the Kekul\'{e} vortex and they move around lattice together with
    the vortex. The location of the vortex is controlled by gate
    voltages on the nanowires. Here the vortex has charge $q = -
    1$. All other parameters are the same as in
    Fig. \ref{fig:majorana-lattice}.}
  \label{fig:vortex}
\end{figure*}

In Fig.~\ref{fig:vortex}a,c we show the electronic LDOS at zero energy
for the wire network in Fig.~\ref{fig:brickwall} with a Kekul\'{e}
vortex pattern of applied gate voltages. In Fig.~\ref{fig:vortex}b,d
we plot the intensity of eigenfunctions associated to the zero modes
bound to the Kekul\'{e} vortex. We choose the vorticity to be
$-1$. Other parameters are the same as those in
Fig. \ref{fig:gappedlattice}

The results described above establish that the wire network presented
in Fig.~\ref{fig:brickwall} is a concrete realization of the proposal
in Ref. \cite{yangHierarchicalMajoranasProgrammable2019a} to obtain
logical Majorana zero modes in a hierarchical manner. The architecture
in Fig.~\ref{fig:brickwall} enables the placement of multiple vortices
and the movement of those vortices by simply changing the gate voltage
on the wires according to Eq.~(\ref{eq:vortex}). In particular, this
construction allows the logical Majorana zero modes to be braided
adiabatically by the modulation of the gate voltages.

\section{Connection to experimental setups}
\label{sec:parameters}

We now connect the idealized tight-binding model used to describe the
Majorana network with a more realistic model of semiconductor system
proximitized with s-wave superconductors. We return to the Hamiltonian
in Eq. (\ref{eq:Hcontinuum}) and consider an infinite nanowire in the
momentum space representation, yielding
\begin{equation}
  H_{\rm wire} = \frac{1}{2} \sum_k \psi_k^{\dagger} H_k \psi_k,
\end{equation}
where
\begin{equation}
  H_k = (\varepsilon_k + \lambda k \sigma_y - \mu_w)
    \tau_z + E_Z \sigma_z + \Delta_s \tau_x,
\end{equation}
$\varepsilon_k = \hbar^2 k^2 / 2 m^\ast$, and $E_Z = g \mu_B | B
|$/2. The eigenvalues of this matrix are
\begin{equation}
  E_k = \pm \sqrt{(\varepsilon_k - \mu)^2 + E_Z^2 + \Delta_s^2 +
    \lambda^2 k^2 \pm 2 R_k},
  \label{eq:eigenvalue}
\end{equation}
where
\begin{equation}
R_k = \sqrt{(\varepsilon_k - \mu)^2 (E_Z^2 + \lambda^2 k^2) +
  \Delta_s^2 E_Z^2}.
\end{equation}
Each one of the four eigenvalues generates a band in $k$-space. The
exact shape of these bands depends sensitively on the values of the
parameters $m^\ast$, $E_Z$, $\Delta_s$, $\lambda$, and
$\mu_w$. Therefore, it is fundamental to seek parameter values that
match experimental systems. For that purpose, we choose InSb-NbTiN
hybrid nanowires, which are currently used to realize Majorana zero
modes. They have a proximity-effect induced superconductor gap
$\Delta \approx 1$ meV. The effective mass of bulk InSb
is $m^\ast = 0.014\, m_e$, where $m_e$ is the electron bare mass
\cite{Nilsson2009}. The Rashba spin-orbit coupling parameter for
bulk InSb is $\lambda = 0.1$~eV$\cdot$nm and the g-factor is $50$
{\cite{lutchynMajoranaZeroModes2018}}. Since it is advantageous to use
a large magnetic field and the critical field for bulk NbTiN is
approximately 10~T, we pick this value for our analysis. Thus,
following Eq. (\ref{eq:critical_B}), the range of chemical potential
values for which the nanowire remains in the topological phase is
$|\mu_w| \lesssim 11$~meV.

After substituting those experimental parameter values into
Eq. (\ref{eq:eigenvalue}) we find low-lying energy bands which can we
well approximated by the dispersion relation
\begin{equation}
E_k \approx \sqrt{\alpha (k\pm k_0)^2 + \beta},
\label{eq:E_approx}
\end{equation}
with $\alpha \approx 0.140$~eV$^2\cdot$nm$^2$, $k_0\approx
0.0777$~eV$^2\cdot$nm, and $\beta \approx 2.23\times 10^{-7}$~eV$^2$.

We can similarly derive a low-lying band structure from the Kitaev
chain Hamiltonian in Eq. (\ref{eq:HKitaevTRS}). In the long
wave-length limit, we find an expression that matches
Eq. (\ref{eq:E_approx}), allowing us to connect its coefficients with
the Kitaev chain parameters as follows:
\begin{equation}
  t\, a^2 = \frac{\sqrt{\alpha}}{k_0},
\end{equation}
\begin{equation}
  t - \frac{\mu}{2} - \frac{\Delta^2}{t} = \frac{\sqrt{\alpha}
    k_0}{2},
\end{equation}
and
\begin{equation}
\beta = \frac{\Delta^2}{2t} \left[ 2t - \mu -
  \frac{2\Delta^2}{t}\right],
\label{eq:beta}
\end{equation}
where $a$ is the chain lattice constant. These relations are obtained
under the assumption that $\mu < 2t - \Delta^2/t$. Inserting the
experimental values for $\alpha$, $k_0$, and $\beta$ into these
relations, we find that a realistic Kitaev model parameters satisfy
\begin{equation}
  t\, a^2 \approx 4.8\ {\rm eV} \cdot{\rm nm^2}
  \label{eq:a-t}
\end{equation}
and
\begin{equation}
t - \frac{\mu}{2} - \frac{\Delta^2}{t} = 0.015\,{\rm eV}.
  \label{eq:Delta-exp}
\end{equation}
Let us consider the case when $\Delta = 0.5t$ and $\mu=0.5t$,
corresponding to the regime of Figs. \ref{fig:Y-junction} and
\ref{fig:5-wire-junction}. Substituting these values into the above
equations we find $t=7.3$~meV and $a=26$~nm; the latter value, when
combined with a length of 20 sites ($L=20$) yields a wire of
approximately $500$~nm in length, which is quite reasonable when
considering a realistic nanowire. Equation (\ref{eq:beta}) serves as a
consistency check: The r.h.s. yields 1~meV, which is about 5 times
larger than the fitted value for $\beta$. This discrepancy comes
primarily from $\Delta$, which is set to a relatively high value in
the numerical calculations to keep the Majorana zero modes
sufficiently isolated at the ends of the chains. Smaller values of
$\Delta$ could be implemented at the expense of using longer chains
(i.e., more sites) and performing exact diagonalizing of larger
systems. However, given that the values obtained for $a$ and $t$ in
comparison to the realistic nanowire model are reasonable, and they
dependent only weakly on $\Delta$ when $\Delta \ll t$, our
considerations show that it is possible to achieve the necessary
conditions for the realization of Majorana zero modes in current
experimental setups.



\begin{figure}[h]
  \raisebox{0.0\height}{\includegraphics[width=9cm,height=6cm]{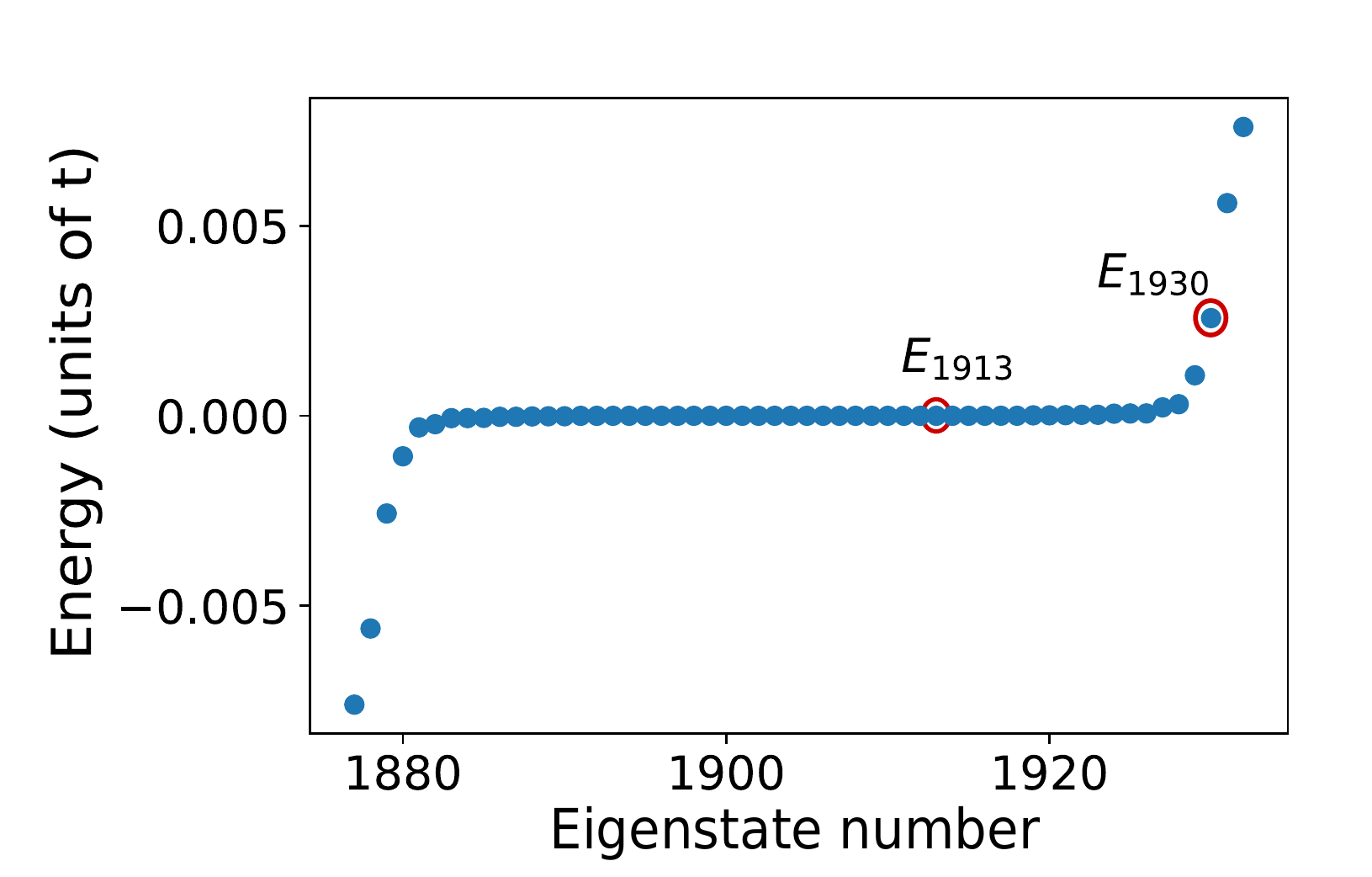}}
  \caption{Energy eigenstates for the brickwall lattice in the
    presence of a vortex. The $E_{1913}$ eigenstate along with its
    particle-hole partner is the vortex eigenstate. The closest state
    which does not contribute to the boundary zero modes is the state
    $E_{1930}$.}
  \label{fig:vortexstates}
\end{figure}

Another important aspect to consider in connection to the experimental
observation of Majorana zero modes is the necessary energy
resolution. In Fig. \ref{fig:vortexstates}, we show a portion of the
energy spectrum for the brickwall lattice for the specific vortex
position in Fig.~\ref{fig:vortex}a. We notice that the difference in
energy between the nearest bulk zero mode state is 0.00257$t$. Using
the value of $t$ obtained from the fitting to the realistic nanowire
model, this energy separation equals approximately 19~$\mu$eV, or,
equivalently, 220~mK, which is a very accessible temperature.

\section{Summary}
\label{sec:conclusions}

In this work we present a nanowire architecture, shown in
Fig.~\ref{fig:brickwall}, where it is possible to realize logical
Majorana zero modes that are movable in 2D by changing gate voltages
on the nanowires. This architecture realizes the hierarchical
construction of
Ref.~\cite{yangHierarchicalMajoranasProgrammable2019a}, without the
need for breaking time-reversal symmetry (TRS).

The basis for building the logical Majorana zero modes is a
programmable ``Majorana graphene'' platform, where a single Majorana
quasi-zero mode at each site of a brickwall or of a honeycomb lattice
hybridizes with modes on the three neighboring sites. The degree of
hybridization, and hence the effective hopping, is controlled by gate
voltages. To arrive at the geometry in Fig.~\ref{fig:brickwall}, we
showed that junctions of five nanowires meeting at each site are
necessary so that (i) there is a single (quasi-)zero mode in each
site; and (ii) this single mode hybridizes with the three neighboring
sites. The number of nanowires needed at the junction follows from two
indices constructed from the polarities of the zero modes at the end
of nanowires. (A positive polarity corresponds to zero modes that are
even under TRS, while a negative polarity corresponds to modes that
are odd under TRS.) For a junction where $n_+$ positive and $n_-$
negative polarities meet, the number of zero modes at the junction is
$\nu=|n_+-n_-|$ and the wave function of the zero modes spread over
$\rho=\max(n_+,n_-)$ wires. We thus satisfy conditions (i) and (ii)
with either $n_+=3$ and $n_-=2$, or $n_+=2$ and $n_-=3$, which are the
cases in the two sublattices of the brickwall network shown in
Fig.~\ref{fig:brickwall}. In the paper, we show numerical results
obtained from electronic tight-binding models of nanowire junctions
are in agreement with this counting.

We further carried out numerical studies of a tight-binding model for
all nanowires and junctions of the network in
Fig.~\ref{fig:brickwall}. In particular, we decorated the
tight-binding model with a Kekul\'{e} dimerization pattern and showed
that it is possible to make the bulk of the system gapped. We achieved
the final stage in the hierarchical construction of
Ref.~\cite{yangHierarchicalMajoranasProgrammable2019a} by including
vortices in the Kekul\'{e} dimerization pattern and showed that there
exists a zero mode -- the logical Majorana mode -- at the core of the
Kekul\'{e} vortex. Finally, we provided estimates of the experimental
values for the parameters used in the numerical calculations and
argued that it is possible to detect the logical Majorana zero modes
using low-temperature local probes.

In closing, we stress that the construction of movable logical
Majorana zero modes in 2D would enable direct and controllable
experiments where Majoranas are braided. This realization of braiding
my require less stringent conditions on the nanowires than other
proposals in the literature.


\begin{acknowledgments}

  The authors would like to thank Hongji Yu for valuable discussions
  at the early stages of this work. This work was partially supported
  by the U.S. Department of Energy, Office of Science, Basic Energy
  Sciences, under Award DE-SC0019275, and by NSF Grant DMR-1906325
  (C.C. and G.G.).

\end{acknowledgments}



\bibliography{Majorana_network_manuscript.bib}





\end{document}